# A Virtual Young's Double Slit Experiment for Hard X-ray Photons

A.F. Isakovic[1], A. Stein[1], J.B. Warren[1], A.R. Sandy[2], S. Narayanan[2], M. Sprung[2*], J.M. Ablett[1#], D.P. Siddons[1], M. Metzler[3], and K. Evans-Lutterodt[1]

[1] Brookhaven National Laboratory, Upton, NY 11973

[2] Argonne National Laboratory, Advanced Photon Source, Argonne IL, 60439

[3] Cornell NanoScale Science & Technology Facility, Ithaca, NY, 14850

## Abstract

We have implemented a virtual Young's double slit experiment for hard X-ray photons with micro-fabricated bi-prisms. We observe fringe patterns with a scintillator, and quantify interferograms by detecting X-ray fluorescence from a scanned 30nm Cr metal film. The observed intensities are best modeled with a near-field, Fresnel analysis. The maximum fringe number in the overlap region is proportional to the ratio of real to imaginary parts refractive index of the prism material. The horizontal and vertical transverse coherence lengths at beamline APS 8-ID are measured.

41.50.+h, 07.85.Tt , 07.85-m, 42.25. Hz

The prism as an optical element for hard x-ray photons was introduced over 70 years ago, for example in attempts to determine the ratio **e/m** [1]. However, the limited useful aperture results in hard X-ray prisms (HXP) being seldom used, in contrast with the many applications of visible light prisms. The limited aperture results from the small deflection angles possible with refraction at a single interface, combined with a relatively large absorption coefficient. Based on the wide array of uses for visible light prisms, it is clear that HXP's could have a significant impact in a variety of fields such as X-ray interferometry, holography and phase contrast microscopy. We have fabricated arrays of silicon bi-prisms with standard micro-fabrication methods, and have used them to implement interferometers with which we have measured the transverse coherence lengths of beamline 8-ID at APS.

Flexible ways to obtain interference patterns sensitive to the transverse coherence length for hard X-ray photons are of great current interest [2-6], but there have been relatively few successful examples. A natural way to obtain interference patterns is to perform a Young's double slit (YDS) experiment. In this approach, a pair of pinholes is illuminated by a beam, where the visibility of the interference fringes can be correlated to the transverse coherence of the beam. Although conceptually simple, it is difficult to implement the ideal YDS experiment for photons in the hard X-ray spectrum. The pinholes must be small ($\sim 10^{-6}$ m), yet the film needs to be sufficiently thick ($\sim 10^{-4}$ m) to attenuate all photons not passing through the pinholes. The resulting large aspect ratio of film thickness to pinhole diameter ($\sim 10^{-4}$m/$10^{-6}$m = $10^2$) is difficult to fabricate. Even if fabrication is successful, the aspect ratio of the pinhole [7] is such that the pair of pinholes may be more accurately modeled as two weakly coupled waveguides, a more complicated system to analyze than simple YDS model. In spite of these difficulties, the pinhole YDS experiment was employed successfully at lower photon energies [8,9]. However, as the experiment [9] showed, at the higher energies, the film becomes more transparent, and the classic YDS pattern is overwhelmed by interference between each pinhole separately with the transmission through the film.

We chose instead to implement a virtual YDS by micro-fabricating bi-prisms, examples of which are shown in Fig.1 (a, b). In this approach, light is deflected by an angle $\theta_T$, by a pair of prisms to a region of overlap where a fringe pattern is generated, as depicted in Fig. 1(c). The refracting surface has an angle $\theta_P$ with respect to x-axis. The effect of each prism is to create a separate virtual source, and one can view the resulting fringe field as originating from the interference due to the two virtual sources. In this wavefront-division method, the transverse coherence length $\xi$ sets an upper bound on the useful size of the prism that produces interference fringes. The number of fringes at the maximal overlap distance L from the prism $L \cong \xi/2\theta_T$ is given by $N \cong \xi/(\lambda/2\theta_T)$. In order to get at least one fringe, one finds that $\theta_T > \lambda/2\xi$. This implies that $L < \xi^2/\lambda$, and consequently all fringes obtained with this bi-prism method are going to be in the near field, Fresnel regime, not the far field Fraunhofer regime, as is the case for the conventional YDS.

The maximum number of fringes is constrained by the absorption in the prism as we now show. The deflection due to a single prism is given by $\theta_T \approx \delta \tan(\theta_P)$. For a transverse coherence length $\xi$, the maximum useful $\theta_P$ of the prism is set by $\tan(\theta_P) = L_{ABS}/(\xi/2)$, where $L_{ABS}$ is the absorption length and is $\lambda/(4\pi\beta)$. Thus $\tan(\theta_P) < \lambda/(2\pi\beta\xi)$ and one can show that the number of fringes in the maximal overlap region $N_{max} < \delta/(4\pi\beta)$, and this is dependent only on the ratio of the prism materials real and imaginary parts of the refractive index. Tabulated Si values for δ and β give between 1 and 15 fringes for energies between 5 and 15 keV, which we show is in a very good agreement with the interference patterns presented here. Outside this overlap region fringes are strongly attenuated.

We micro-fabricated bi-prisms in silicon (Si), with angles that range from 25 degrees to 80 degrees, with details discussed in [11,12], except that a 30nm Cr film is used in place of $SiO_2$ film. Measurements were performed at APS beamline 8-ID-I at and relevant details of the beamline are documented in Ref [10]. Some measurements were also performed at NSLS X13B beamline. For imaging purposes, the X-rays hit a YAG-crystal (yttrium-iron-garnet), and the visible light fluorescence is detected by a CCD camera. The insets in Fig. 2 show typical CCD images, but these images are not readily quantified. Instead, we directly measure the fringes with X-ray fluorescence (XRF) from a metal strip. A Si wall, several hundred microns long and about 40 microns high, was micro-fabricated, and capped with a 30-nm thick layer of chromium (Cr) film that serves as a knife-edge (Fig. 1d). An "xyz$\theta_x\theta_y$" micro-translating stage controls the positioning of a small Si wafer bearing this knife-edge, and the "VORTEX" detector [13] is sited about 8 cm away from the knife-edge, in a near-perpendicular orientation to the direction of propagation of the X-ray beam. The Cr strip was oriented with respect to the fringes to maximize the resolution. Slits upstream of the prism restrict light to illuminate the active prism areas, typically 200 microns by 40 microns. One can normalize the data by measuring the field intensity present after translating the prism out of the path.

We show the fringes as a function of three variables, the prism angle in Fig. 2, the photon energy in Fig. 3(a-c), and distance from prism in Fig. 3(d-f). Besides observation of a fringe intensity profile with the expected fringe period given by $\lambda/2\theta_T$, what is immediately striking about the data in Figs. 2 and 3 is the non-monotonic behavior of the intensities of the fringes, which we address in the next paragraph. Some other aspects of the fringe pattern are more readily interpreted, including the dependence on the prism angle and the energy dependency. For example, at a fixed distance of the Cr strip from the bi-prism, larger prism angles $\theta_P$ result in larger deflection angles $\theta_F \approx \delta \tan(\theta_P)$, and this gives a larger region for fringe overlap, and a smaller period $L_P$ for the fringes, where $L_P = \lambda/(2\theta_T)$, as shown in Fig. 2. In Fig. 3 (a-c) we show intensities as a function of photon energy for fixed prism angle (65 deg), illustrating that the deflection angle $\theta_T$ and the overlap region decrease with increasing energy, while $L_P$ is increasing with increasing energy. Conventional optical YDS experiments have a smoothly varying fringe intensity envelope function. The lack of smoothly varying envelope here is a direct result of being in the Fresnel regime. To accurately describe this we have generated a non-linear least squares fitting procedure based on the Fresnel-Kirchoff diffraction formula. If one makes the usual approximation of a "thin" optic [14], one has an equation of the form: $U(x,z) = \frac{1}{i\lambda} \int T(\eta) \frac{\exp(ikr_{01})}{r_{01}} d\eta$, where $k = \frac{2\pi}{\lambda}$, and $r_{01} = \sqrt{z^2 + (x-\eta)^2}$, with ($x$, $z$) defining the image plane at a distance $z$ from the prism and the variable $\eta$ is in the "thin" object plane. We model the prisms as a thin optic with a transmission function with phase shift and amplitude given by $T(\eta) = \exp(-(2\pi\beta*\mathbf{t}(\eta))/\lambda)\exp(i*(2\pi\delta*\mathbf{t}(\eta))/\lambda)$, where $\mathbf{t}(\eta) = \text{abs}(\eta)\tan(\theta_p)$ represents the thickness of the prism, and $\eta$ is the distance from the optical axis.

One can qualitatively understand the unexpected envelope of intensities. If one illuminates either prism of the bi-prism pair separately, one obtains the familiar modulated intensity pattern due to coherent illumination of an aperture [14]. The illumination of both prisms of the bi-prism pair results in the interference fringes, but these fringe intensities are modified by the product of the aforementioned modulation functions. The resulting fringe intensities thus depend on the distance from the bi-prism; and this is most clearly visible in the experimental data presented in Fig. 3(d-f), where the fringe intensities are shown at different distances from the prism. The quality of the overlap between the model and raw XRF data in Fig. 3 further supports this picture.

Given this demonstration of a practical hard x-ray interferometer, a natural experiment is to attempt to determine the transverse coherence length in a beamline, which in the ideal case is due only to the source size and distance from the source. In practice, a beamline often has imperfect optical components that will degrade the beam coherence. At the APS 8-ID beamline the "one sigma" source sizes are quoted [15] to be 11 microns in the vertical and 110 microns in the horizontal. Fig. 4 (a,b) show the resulting fringe intensity profile in the vertical and horizontal directions. One can see immediately that the vertical orientation displays greater fringe visibility and contrast than does the horizontal profile, showing qualitatively that the source size in the vertical is smaller than in the horizontal. In Fig. 4(c) the extracted non-uniform fringe visibility emphasizes the near-field behavior. The maximum fringe visibilities in the vertical are above 0.9, and are about ~0.6 in the horizontal.

We extract the asymmetric source size from the fit to the data-points using our model described above, where the source is modeled as a completely incoherent source of a finite size. From the fits we extract a vertical source size of 10+/-5 microns and a horizontal source size of 130+/-20 microns. These measured source sizes are comparable to the quoted source sizes above, implying that the beamline preserves the coherence of the source quite well in both directions, but is slightly worse in the horizontal direction.

In the YDS, the fringe visibility is uniform and is directly related to the complex coherence factor $\mu_{12}$ [16]. In order for the fringe visibility to be equal to $\mu_{12}$, the intensity from each aperture should be equal, which is not the case here. Symmetry suggests that the central fringe is likely to have equal amplitude from both prisms independent of any details, and should allow its use as a semi-quantitative guide to make comparisons.

We list some of the advantages of using micro-fabricated bi-prisms for an X-ray interferometer, in combination with the XRF method for quantifying the interference fringe intensities. The simplicity of the system allows one to model the measured fringe intensities from first principles. The bi-prism can function as interferometers over a wide range of energies, by simply increasing the detector distance to compensate for the reduced deflection angles at higher energies. The lithographic methods for defining the bi-prisms allow a phase accuracy of order $(50nm/30\mu m)*2\pi \cong 5 \times 10^{-3} \pi$ due to 50nm placement accuracy. Imperfections can be quantified with an SEM and included in the modeling of the system, as modifications to $T(\eta)$. The useful signal is not limited to the flux that passes through sub-micron pinholes, and consequently the VYDS will have improved S/N ratio over YDS. Using the 30nm thin metal film as a detector allows one to quantify the fringe intensities with fringes periods as small as 60 nm; thinner metal films should allow detection of fringe periods down to the 5 nm range. This interferometer is more sensitive to the transverse rather than the longitudinal coherence length. The bandwidth of the monochromator at 8-ID is of order 2 eV, giving a longitudinal coherence of ~0.5 micron. For the typical deflection angles ~$10^{-5}$, a longitudinal coherence length of 0.5 micron will not affect the data shown here.

Real sources may be partially spatially coherent, and a more complete description of the source will be the measured complex coherence factor $\mu_{12}$ [16]. For the classical YDS, $\mu_{12}$ can be extracted from the dependence of the fringe visibility as a function of separation between the slits. For this bi-prism interferometer, a direct route to extraction of $\mu_{12}$ can be deduced from an extension of Schell's theorem [17], which applies even in the near field limit, which is the case here.

In conclusion, using micro-fabricated prisms we have implemented a VYDS for hard X-ray photons. We measure the absolute amplitude of the fringes from the XRF signal from a 30nm Cr metal film scanned through the fringe field. Unlike the classical far-field Fraunhofer analysis typically adopted, we found Kirchhoff-Fresnel near-field analysis needs to be applied for the X-ray bi-prism method at hard x-ray wavelengths. The number of fringes in the classical overlap region is comparable to $\delta/(4\pi\beta)$, the ratio of real to imaginary parts of the refractive index of the prism material, showing a path towards better interferometers. We have shown that for typical monochromatic bandwidths, this interferometer is more sensitive to transverse than longitudinal coherence lengths. Finally, we measured the transverse coherence lengths of the APS 8-ID beamline. These results suggest that micro-fabricated prisms will play a larger role in novel hard x-ray optical arrangements for phase contrast methods and other types of interferometeric imaging and characterization.

Use of the NSLS, the CFN and the NSLS-II project at BNL, was supported by the U.S. DOE, Office of Basic Energy Sciences, under Contract No. DE-AC02-98CH10886, and the use of ANL-APS is supported through the DOE contract DE-AC02-06CH11357. This work was performed in part at the Cornell CNF, a member of the NNIN, which is supported by the NSF. AFI and KEL acknowledge early support through BNL LDRD 06-046, P. Takacs (BNL) for early suggestions about the manuscript, and C. C. Kao (BNL) for support. We thank G. Bordonaro and R. Ilic of CNF for timely technical assistance.*Currently at HASYLAB, #Currently at SOLEIL

**References:**
[1] H.E. Stauss, Phys. Rev. **36**, 1101, (1930) (second series).

[2] A. Snigirev, I. Snigireva, V. Kohn, *et al*., Phys. Rev. Lett. **103**, 064801 (2009), and references therein.

[3] W. Cash, A. Shipley, S. Osterman, M. Joy, Nature **407**, 160 (2000).

[4] T. Weitkamp, A. Diaz, C. David, F. Pfeiffer *et al*., Opt. Expr. **13** (16), 6296 (2005), and references therein.

[5] A. Momose, T. Takeda, Y. Itai, A. Yoneyama, *et al*., J. Synchr. Rad. **5**, 1187 (1998).

[6] A. R. Lang, A. P. W. Makepeace, J. Synchrotron Rad. **6**, 59 (1999).

[7] C. Fuhse and T. Salditt, Optics Communications **265**, 140, (2006).

[8] Y. Liu, M. Seminario, F.G. Tomasel, *et al*., Phys. Rev. A. **63**, 033802 (2001).

[9] W. Leitenberger, H. Wendrock, L. Bischoff, T. Weitkamp, J. Synchr. Rad **11**, 190 (2004), and references therein.

[10] A. R. Sandy, K. Evans-Lutterodt, K. Fezzaa, *et al*., SPIE Proceedings v.6705, 67050N-1, ed. by A. M. Khousnary, C. Morawe, S. Goto, San Diego, CA (2007).

[11] A. Stein, K.Evans-Lutterodt, N. Bozovic, *et al*., J. Vac. Sci. Technol. B **26** (1) 122,A Virtual Young's Double Slit Experiment for Hard X-ray Photons v1 BNL-NSLS          9

Figure captions:

Fig. 1 (a) A top view of an 1D array of X-ray prisms, nanofabricated from Si, with varying prism angle in each channel. (b) the second family of prisms used in this study (c) A schematic (not-to-scale) of the experiment, illustrating the creation of a pair of virtual sources from a single real source; (d) SEM image of the Cr-capped Si wall that serves as the "knife edge" for X-ray fluorescence.

Fig.2 (top to bottom) Evolution of the interference pattern with changing prism angle. Left insets are SEM images of prisms, while the right insets are YAG/CCD images of the corresponding interference patterns.

Fig.3 (a-c) The evolution of the interference pattern with X-ray photon energy. The line is only a guide to the eye. (d-f) The evolution of the fringe patterns as a function of the separation between the 65 deg prism from Fig. 1b, and the Cr "knife edge". The red line is a fit, based on the model presented in the text.

Fig. 4 (a) X-ray interference pattern for a vertical orientation of the X-ray bi-prism. The fit, obtained numerically as described in the text, is a solid line; (b) the same type of measurement and the fit for the horizontal orientation. The right insets in (a, b) are CCD interference patterns; (c) Lateral spatial variation of the *visibility* factor for two mutually perpendicular positions of the X-ray bi-prism optics.

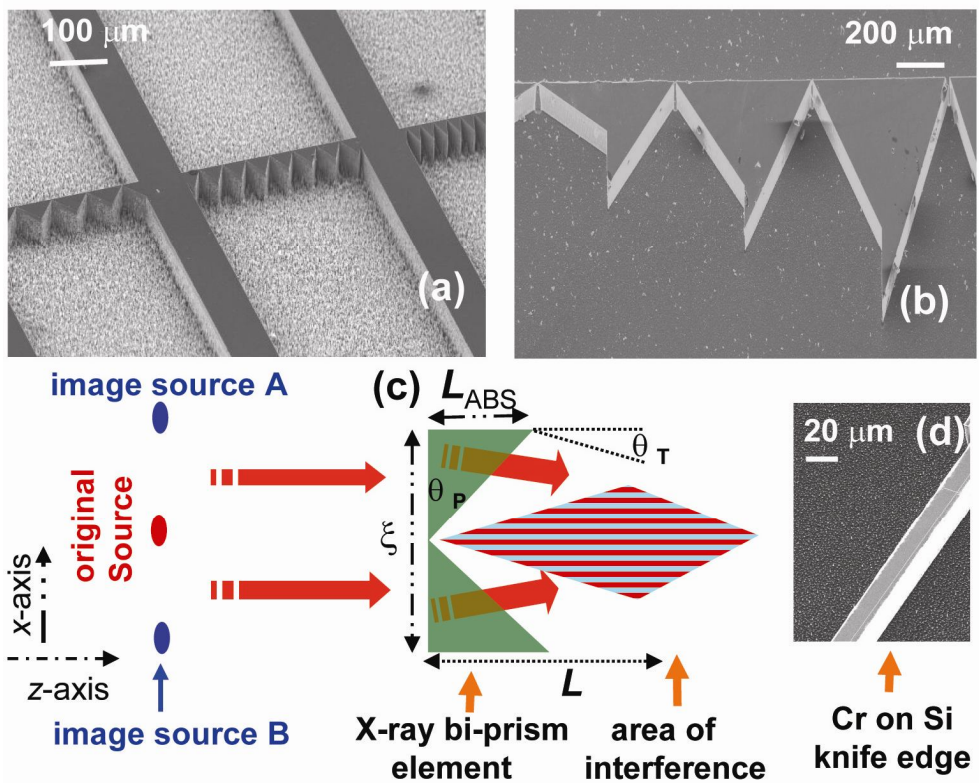

Fig. 1 (a) A top view of an 1D array of X-ray prisms, nanofabricated from Si, with varying prism angle in each channel. (b) the second family of prisms used in this study (c) A schematic (not-to-scale) of the experiment, indicating a role of a single pair of X-ray prisms, and the way a pair acts as a two-point source. (d) SEM image of the Cr-capped Si wall that serves as the "knife edge" for X-ray fluorescence.

"A Virtual Young's Double Slit Experiment for Hard X-ray Photons"
by A. F. Isakovic, K. Evans-Lutterodt *et al*.

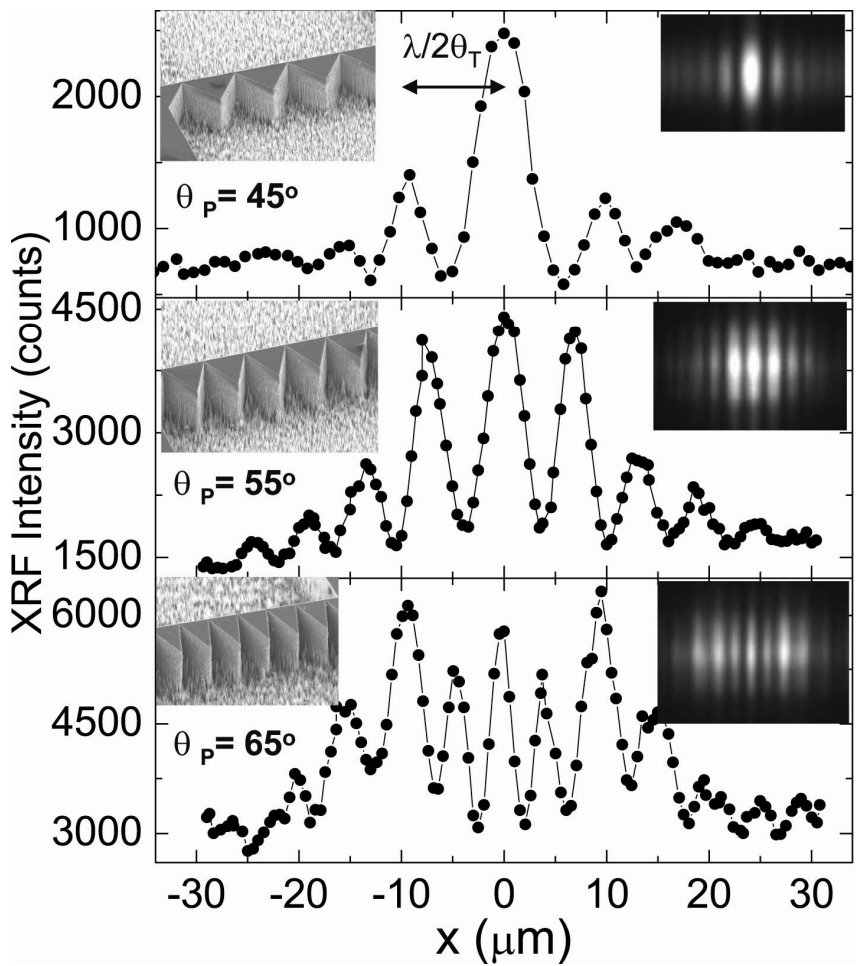

Fig.2 (top to bottom) Evolution of the interference pattern with changing prism angle. Left insets are SEM images of prisms, while the right insets are CCD images of the corresponding interference patterns.

"A Virtual Young's Double Slit Experiment for Hard X-ray Photons"
by A. F. Isakovic, K. Evans-Lutterodt *et al*.

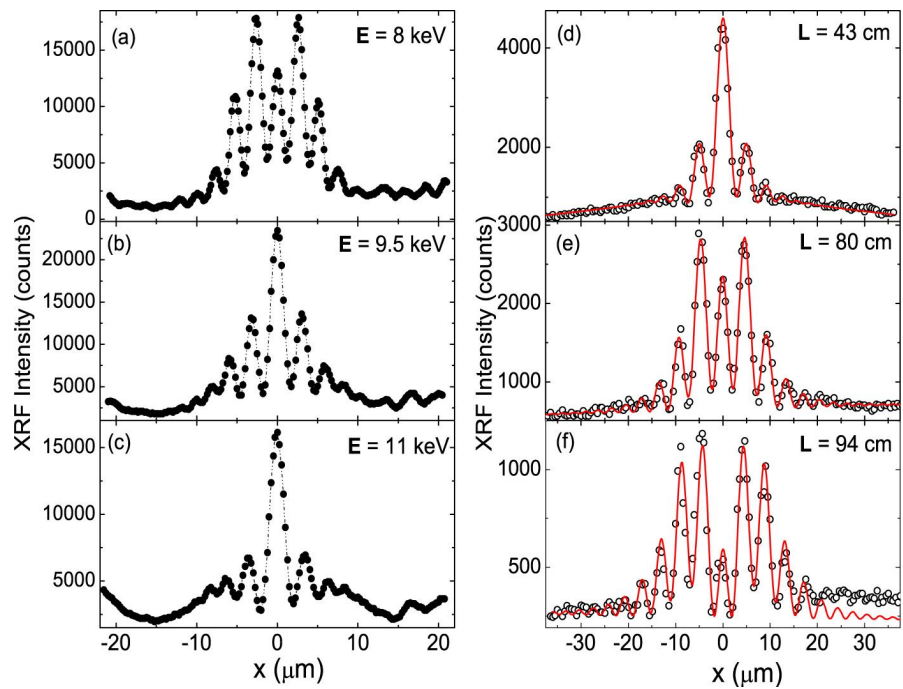

Fig.3 (a-c) The evolution of the interference pattern with X-ray photon energy. The line is only a guide to the eye. (d-f) The evolution of the fringe patterns as a function of the separation between the 65 deg prism from Fig. 1b, and the Cr "knife edge". The red line is a fit, based on the model presented in the text.

"A Virtual Young's Double Slit Experiment for Hard X-ray Photons"
by A. F. Isakovic, K. Evans-Lutterodt *et al*.

A Virtual Young's Double Slit Experiment for Hard X-ray Photons v1 BNL-NSLS        14

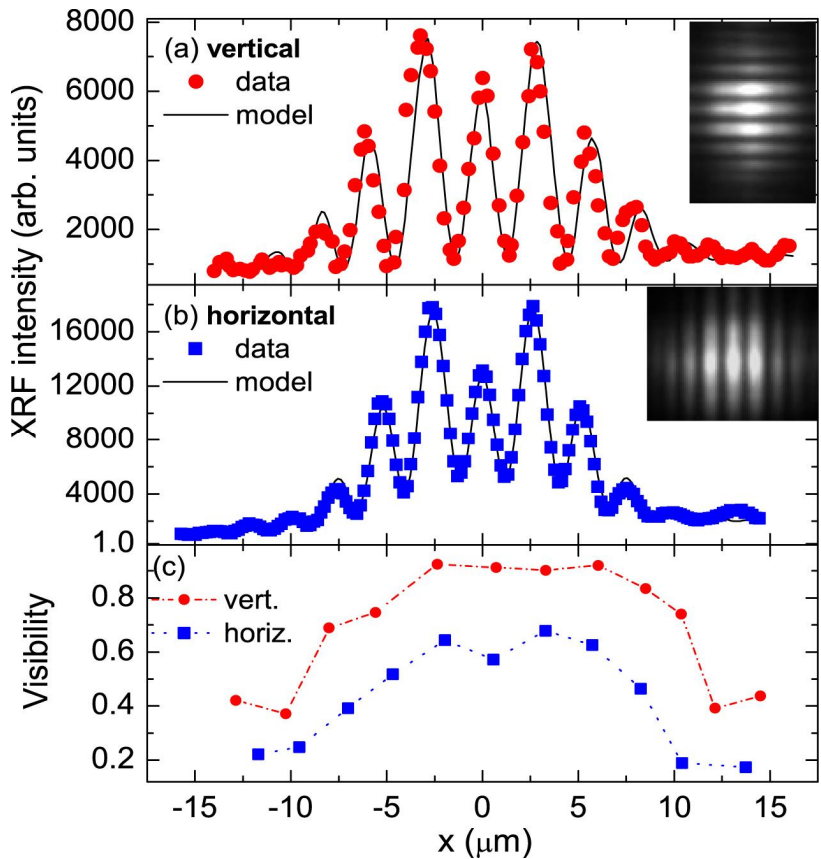

Fig. 4 (a) X-ray interference pattern for a vertical orientation of the X-ray bi-prism. The fit, obtained numerically as described in the text, is a solid line; (b) the same type of measurement and the fit for the horizontal orientation. The right insets in (a, b) are CCD interference patterns; (c) Lateral spatial variation of the *visibility* factor for two mutually perpendicular positions of the X-ray bi-prism optics.

"A Virtual Young's Double Slit Experiment for Hard X-ray Photons"
by A. F. Isakovic, K. Evans-Lutterodt *et al*.